# Local Temperature Redistribution and Structural Transition During Joule-Heating-Driven Conductance Switching in VO$_2$

*Suhas Kumar*[1,2], *Matthew D. Pickett*[1]*, *John Paul Strachan*[1], *Gary Gibson*[1], *Yoshio Nishi*[2], and *R. Stanley Williams*[1]

[1]*Hewlett-Packard Laboratories, 1501 Page Mill Rd, Palo Alto, CA 94304, USA*
[2]*Stanford University, Stanford, CA 94305, USA*
*E-mail: Matthew.Pickett@HP.com*

**Joule-heating induced conductance-switching is studied in VO$_2$**, a Mott insulator. Complementary *in-situ* techniques including optical characterization, blackbody microscopy, scanning transmission x-ray microscopy (STXM) and numerical simulations are used. Abrupt redistribution in local temperature is shown to occur upon conductance-switching along with a structural phase transition, at the same current.

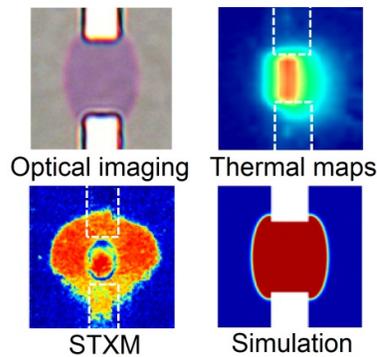

Vanadium dioxide is one of many correlated-electron transition metal oxides that undergo an insulator-metal transition (IMT) driven by various stimuli including temperature, strain, doping and optical excitation.[1] More specifically, VO$_2$ undergoes both an IMT and a structural phase transition (SPT) from a semiconducting monoclinic to a metallic rutile phase at a temperature of roughly 340 K.[2,3] IMT materials interposed between two metallic electrodes are known to exhibit volatile conductance switching,[4] while these materials have applications ranging from non-volatile memory[5] to neuromorphic computing.[4] In such two-terminal devices, the driving mechanism behind the IMT-induced conductance-switching has been under debate, with competing claims that an electric-field[6-8] or Joule-heating[9] is the dominant process. There is also debate about whether or not the SPT is triggered concurrently with the electrically induced IMT.[10,11] Here we use a combination of optical, infrared and x-ray spectro-microscopy techniques to show that driving the device beyond the switching threshold results in an abrupt redistribution of local temperature as well as the emergence of metallic filaments with rutile phase ordering. We additionally show that the thermal redistribution and the SPT occur at the same applied current even if the SPT occurs a few degrees kelvin after IMT, consistent with a Joule-heating driven phase transition model. These results demonstrate that spatially resolved techniques are critical for unambiguously characterizing and understanding this inhomogeneous phenomenon.

From both technological and scientific standpoints it is important to unambiguously understand the driving mechanism behind conductance switching in VO$_2$, especially since the material exhibits both an IMT and an SPT which can be triggered by a range of extrinsic forces.[2,10,12,13] When switching was first observed in VO$_2$, it was believed that Joule-heating caused a temperature-driven IMT simultaneously with the SPT, giving rise to the current-controlled negative differential resistance (CC-NDR) and conductance-switching.[14-16] This conclusion has been challenged by recent claims of electric-field-driven switching in which Joule-heating was negligible.[6,17] To add additional complexity to the understanding, it remained unclear as to whether or not SPT occurs, even under a Joule-heating model, since $T_{SPT}$ can be higher than $T_{IMT}$.[10,11] Here we employ spatially-resolved observations of the local temperature and electronic structure of planar VO$_2$ devices in order to directly characterize the inhomogeneous behavior of the device during switching. Using these techniques we observed that the sharp conductance-switching phenomenon coincided with the emergence of filamentary optical contrast between electrodes, a sharp redistribution of the thermal profile concentrated around the filament, and the emergence of filamentary rutile phase-ordering in electronic structure. The coincidence of these phenomena at the same applied current is described well by a Joule-heating driven thermal phase transition model in which both the IMT and SPT occur at or near the same temperature. Some of the individual techniques discussed here have been applied to similar devices, but without the spatial resolution required to observe redistributions in both the local temperature and electronic structure of a functioning device.[8,12,16,18]

In order to characterize the conductance switching phenomenon in VO$_2$, we fabricated planar devices by evaporating two platinum electrodes of width W separated by a gap of length L onto a thin VO$_2$ film (Figure 1a). Data reported in Figures 1-3 were obtained from the same device, with L,W = 16,8 µm. These devices exhibited a sharp drop in low voltage (<50 mV) resistance, $R_0$, as the ambient temperature, $T_{amb}$, was raised above insulator to metal transition temperature ($T_{IMT}$ = 340 K) and exhibited hysteresis upon cooling (Figure 1a). When larger currents (>1 mA) were applied to devices at $T_{amb}$ = 300 K they exhibited switching to a low-resistance state, or turned ON, at threshold point ($I_{ON}$, $V_{ON}$) (Figure 1b). When the current was subsequently ramped down below a second threshold point ($I_{OFF}$, $V_{OFF}$) the device reverted to its original high resistance state, or turned OFF. The switching process results in a two-terminal current-voltage curve which is dynamical, hysteretic, purely dissipative, and reversible and can be mathematically described under a memristive system model.[19,20] The threshold positions depended on ambient temperature, as evidenced by measurements of $V_{ON}$ which monotonically decreased with increasing temperature and approached zero as $T_{amb}$ approached $T_{IMT}$ (~340 K) (Figure 1c). The temperature dependence of $R_0$ and $V_{ON}$ are consistent with previously reports for similar devices.[15,21]

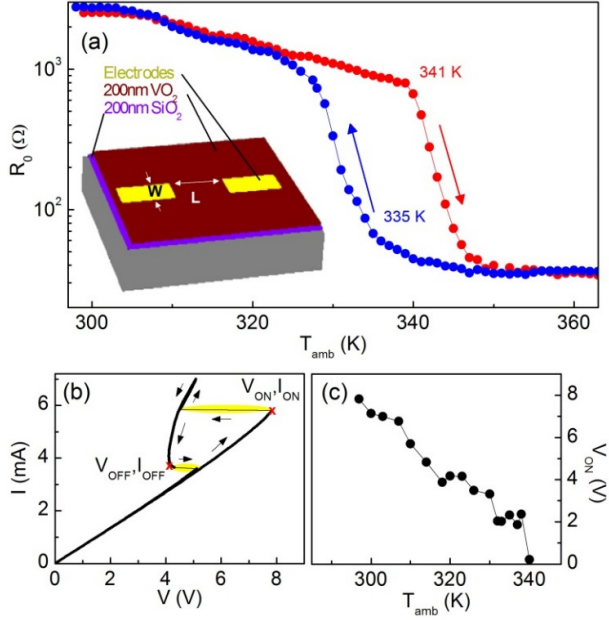

**Figure 1:** (a) Low bias resistance ($R_0$) of the VO$_2$ film showing hysteretic behavior as the ambient temperature ($T_{amb}$) is cycled. Temperatures marked are the transition temperatures for IMT hysteresis on the heating ($T_{IMT}$) and cooling ($T_{IMT}'$) branches. A schematic of the device geometry is included as an inset where 'L and 'W' indicate the length of the gap and width of the electrodes, respectively. (b) Current-voltage curve of a typical planar VO$_2$ device exhibiting the characteristic switching behavior. The ON and OFF thresholds are marked with red 'x' while the NDR regions are highlighted in yellow. (c) The ON-switching voltage ($V_{ON}$) as a function of the ambient temperature of the film.

In the low resistance state, e.g. when currents above $I_{ON}$ were applied, a visible filament emerged which bridged the electrodes. This filament was present only while the device was held in the low resistance state, grew in width with increasing current (Figures 2a-2b), and disappeared when the applied current was reduced below $I_{OFF}$. Optically visible filaments have been previously reported in similar planar VO$_2$ devices and were previously attributed to a local temperature increase within the filaments[22]. To confirm the nature of the optical contrast in our devices, we measured the thermo-reflectance of the VO$_2$ film, without applying current, as a function of ambient temperature (Figure 2c). The reflectance at green wavelengths (~530 nm) decreased with temperature until it saturated at around 353 K, consistent with previously reported data,[23] indicating that the optical contrast is likely related to localized temperature changes. A spatially localized reflectance spectrum was measured by focusing the detector near the edge of the filament and revealed a spectrum similar to the spectrum expected for the material at $T_{amb} = T_{IMT}$ (black curve in Figure 2c). This result qualitatively suggests that the high contrast region is at a temperature above $T_{IMT}$ and that the halo around the filament reflects a thermal gradient to ambient temperature outside of which the material remains insulating. Since the thermoreflectance saturates just above $T_{IMT}$, it is not suitable to perform thermometry and hence we used black-body spectro-microscopy.

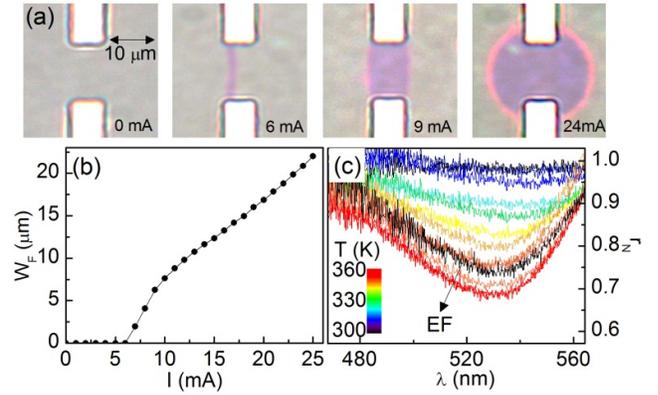

**Figure 2:** (a) Optical micrographs of filaments observed between the electrodes at different applied currents. (b) The variation of the filament width ($W_F$) with current obtained from optical micrographs. (c) VO$_2$ thermo-reflectance spectra ($r_N$) obtained at various uniform ambient temperatures, normalized to the room temperature spectrum. The black curve indicated as 'EF' (electrical filament) was obtained by passing a current of 15 mA through a device and focusing the detector on the edge of the filament region.

In order to quantitatively map local temperature of the device *in situ* throughout the ON and OFF switching processes we employed an InfraScope[24] black-body spectro-microscope. A series of temperature maps, emissivity-calibrated to a temperature resolution of ~ 1 K and with a diffraction-limited spatial resolution of ~ 2 µm, were acquired at current steps of 10 µA. Figure 3a shows four of the temperature maps in which the current was ramped beyond the on-switching threshold, revealing the emergence and growth of a filament-shaped hot spot between the electrodes. We characterized this changing temperature profile by plotting a line scan of the thermal map taken across the central position between the electrodes (hash marks in Figure 3a) as a function of applied current for the ramp up and ramp down (Figure 3b). This figure reveals a discrete reorientation of the thermal profile at the switching currents, as shown in more detail in Figure 3c for the line profiles just before and just after ON and OFF switching. The thermal profile is relatively more concentrated while the device is in the low resistance state and relatively more diffuse while the device is in the high resistance state. We note also that just before ON switching the maximum temperature was measured to be just below the heating IMT temperature while just before OFF switching the maximum temperature was measured to be just above the cooling IMT temperature (both of which were measured previously from Figure 1a). These two observations are displayed quantitatively in Figure 3d in which the distribution width (defined as the width of the region which contains the middle 30% of the area under the temperature distribution) and maximum temperature are plotted as a function of current.

The coincidence of conductance switching with the local crossing of the IMT temperatures clearly demonstrates the thermal nature of switching in these planar devices and the detailed thermal distribution change provides a good description of the switching process. Since the current density, and resulting Joule-heating thermal profile, is inhomogeneous in this device geometry it is only the maximum temperature portion of the device that dictates switching.

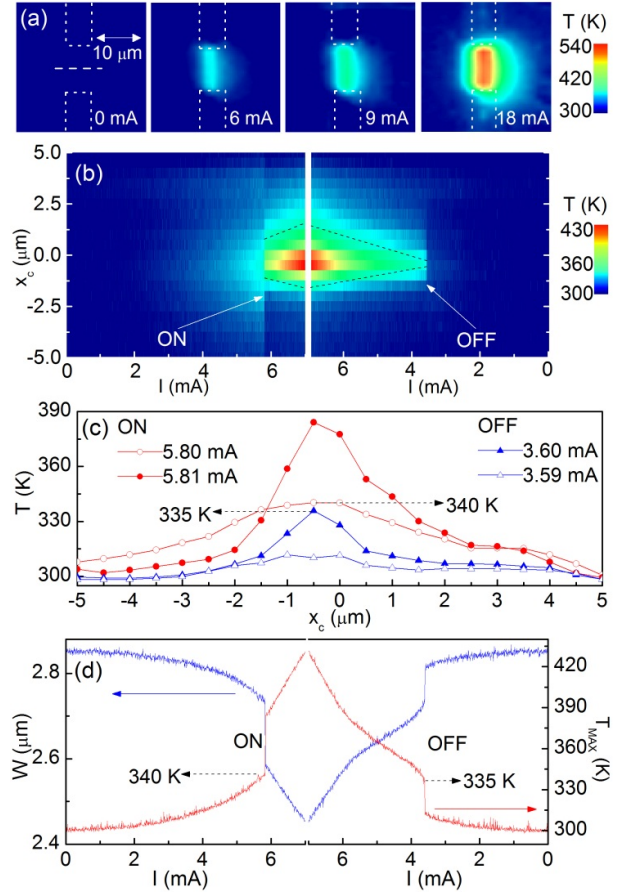

**Figure 3:** (a) Blackbody emission maps of the device at different currents. (b) Temperature profile along a line between the electrodes (indicated as a dashed line in panel (a) at 0 mA) with increasing and decreasing current. $x_c$ is the distance from the center of the electrodes along this line. The discontinuities in temperature at $I_{ON}$ and $I_{OFF}$ are marked. The dotted line is the approximate location of 340 K, the edge of the filament. (c) Temperature profile along the same line used in (b) just above and just below $I_{ON}$ and $I_{OFF}$. Dotted horizontal lines represent $T_{IMT}$ and $T_{IMT}'$ (measured from Figure 1a). (d) Full width, W, at 30% area of the temperature profile along the same line used in (b) for increasing and decreasing currents. The maximum temperature, $T_{MAX}$, along this line is also plotted. Panels (b)-(d) were measured simultaneously with I-V trace of Figure 1b with current steps of 0.01 mA.

Once the switching process is initiated at a point the local resistance drops, causing an unstable positive feedback effect which causes more current to flow into the local low-resistance filamentary pathway until some significantly different steady state is reached. Thus the end result of the switching process is a discontinuous jump in maximum temperature and a coincident focusing of the current into a filament. Interestingly, this redistribution results in a drop in local temperature of around 15 K at some points (e.g. $x_c$ = -2 µm in Figure 3c) despite an increase in current. The maximum temperature displayed in Figure 3d changes more rapidly with current in the on-state compared to the off-state. This is because incremental change in current has a bigger effect on local temperature in a confined filament in the on-state than in a uniform current distribution in the off-state. This observation also counters prior predictions of temperature distributions within a metallic filament.[22] Due to the inhomogeneity of the transition, it is critical to perform high-spatial-resolution thermal measurements since focused single-point or spatially averaged temperature measurements could grossly underestimate the switching temperature leading to an incorrect conclusion on the nature of switching.

In order to directly map structural changes that are associated with inhomogeneous heating, we used scanning transmission x-ray microscopy (STXM) to study the localized electronic structure changes *in situ* during conductance-switching. Since STXM is a transmission technique we fabricated a set of devices on 20 nm thick silicon nitride membranes using an otherwise identical process to the one used for optical and thermal measurements above. Due to the constraints of the two measurement techniques, we fabricated two varieties of devices for characterization: one for black-body spectro-microscopy and another for X-ray measurements. STXM enables x-ray spectroscopic mapping with a 30 nm spatial resolution and 70 meV energy resolution, giving local electronic structure measurements throughout the device.[25] Figure 4a shows two x-ray absorption spectra (XAS) of the oxygen K-edge, collected at a single point midway between the electrodes with no applied current and with current significantly above $I_T$. XAS has been studied extensively in the $VO_2$ system, and the O K-edge was shown to be particularly sensitive to modifications in the structural ordering.[26] The differences observed in the electronic structure of the two spectra, particularly the emergence of the $d_\parallel^*$ band, are attributed to known differences between semiconducting monoclinic (M1) and metallic rutile (R) phase-ordering, showing that there is indeed a SPT.[10,27] We used these point spectra to identify particular absorption energies that gave the maximum contrast between the two phases in STXM, 528.95 eV and 530.2 eV, shown at the bottom of Figure 4a.

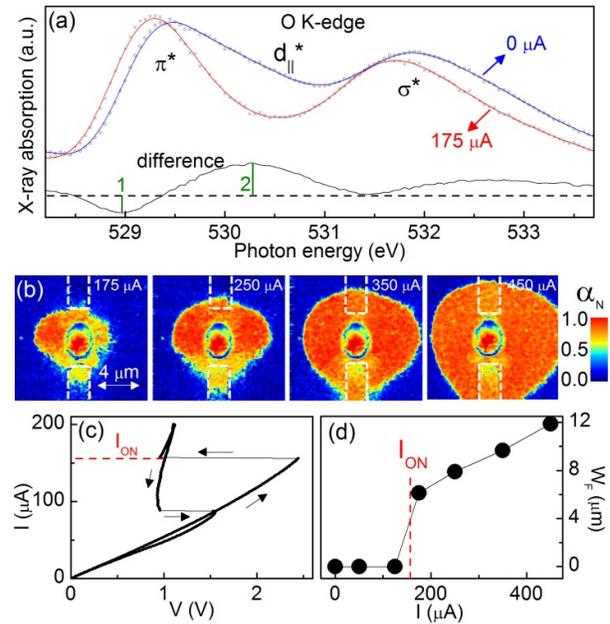

**Figure 4:** (a) X-ray absorption spectra of the region at the midpoint between the device electrodes with applied current above (red) and below (blue) the switching threshold. The above threshold spectrum shows peaks characteristic of R phase ordering while the below threshold spectrum shows peaks characteristic of the M1 phase-ordering. The lower difference curve shows the points of maximum difference '1' (528.95 eV) and '2' (530.2 eV) used to define a normalized phase contrast $\alpha_N$. (b) Maps of phase contrast showing a transformed region between the electrodes that grows with applied current where red indicates a region of with electronic structure having R phase-ordering and blue indicates M1 phase ordering. The circular feature in the center is likely an artifact due to damage introduced by cycling the device through locally elevated temperatures on a 20 nm thin membrane. Current-voltage characteristics (c) and the dependence of the filament width ($W_F$) on current (d) are shown for the same device.

Using these maximum contrast energies we collected spatially resolved absorption maps at both energies for a set of different applied currents and constructed contrast maps, calculated as the normalized ratio of absorption at the two energies, for increasing current (Figure 4b). These maps show regions of high contrast, signaling the presence of metallic rutile phase-ordering, between the electrodes with a similar shape to the thermal hotspots of Figure 3a. As shown in Figures 4c-4d, rutile phase ordering was observed only above the current threshold $I_{ON}$ and grew in width with a similar trend as the thermal hotspots and optical contrast regions. These results show that, although $T_{SPT}$ could be higher than $T_{IMT}$ by several degrees in some cases,[10,11] the discrete jump in local temperature of up to 45 K associated with conductance-switching in this configuration is sufficient to cause a SPT whether or not they actually occur at the same temperature.

We used a 3D finite element COMSOL simulation of our model for thermal switching in order to validate its mathematical feasibility based on fundamental physics. The simulation included basic electrical and thermal physics, including electron drift and heat diffusion, a three-dimensional model of the planar device, and literature values for the materials parameters. To implicitly account for the phase transition of the $VO_2$ without introducing any explicit physics for phase transition into the model, we used simple sigmoid functions for the dependence of its electrical conductivity on temperature. We roughly calibrated this functional form to the data of Figure 1a. Figure 5a shows simulated temperature maps for different currents. For comparison to the optical results of Figure 2a, we clamped the color scale of the simulations to coincide with the saturation temperature of thermo-reflectance, 353 K. The I-V curve and the current dependence of the width of the filaments obtained from simulations are also shown (Figures 5b-5c). Data from simulation qualitatively agree with the experimental data while the quantitative differences could be attributed to several computational parameters that may not account for a real-world scenario. We also note that unlike µm-wide filaments observed here, on-switching might happen along nano-scale percolation paths, depending on the properties of the film.[8] Simulating uniform heating in such a situation could result in a considerable underestimation of local temperature at the switching threshold.[7]

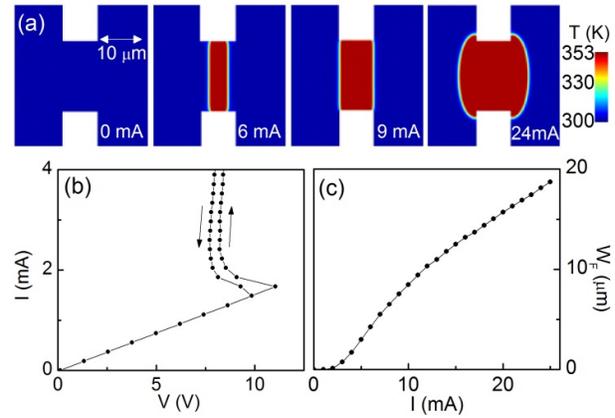

**Figure 5:** (a) Simulated temperature maps at different currents. The scale is clamped at 353 K to account for the saturation of optical contrast at 353 K observed in the thermoreflectance data. I-V curve (b) and the width of the filaments at various currents (c) obtained from simulations.

In conclusion, we have utilized a complementary set of techniques to study the electrical and thermal behavior, and electronic structure of fabricated $VO_2$ devices. Blackbody spectro-microscopy provided a comprehensive description of abrupt redistributions in local temperature during Joule-heating-driven IMT. Filamentary conduction was accounted for in numerical simulations using a simplified electro-thermal model. Using STXM maps we observed that the SPT occurred beyond the same current threshold that induced the IMT although the SPT could occur a few degrees kelvin after the IMT. The occurrence of IMT and SPT at the threshold current was justified by the jump in local temperature upon on-switching. The techniques employed in this study provide a spatially-resolved, direct and unambiguous means to study local temperature and local structure correlated with electronic transport. These results present a clear understanding of the driving mechanism in metal-insulator transition devices, which have gained recent interest due to their promising applications.

**Acknowledgements**
The Advanced Light Source (beamline 11.0.2) at Lawrence Berkeley National Laboratory in Berkeley, CA, USA was used to perform STXM measurements. The Advanced Light Source is supported by the Director, Office of Science, Office of Basic Energy Sciences, of the U.S. Department of Energy under Contract No. DE-AC02-05CH11231.

[1] P. Limelette, A. Georges, D. Jerome, P. Wzietek, P. Metcalf, J. M. Honig, Science **2003**, 302, 89; C. Aurelian, G. Julien, B. Pierre, O. Jean-Christophe, C. Corinne, Advanced Microwave and Millimeter Wave Technologies Semiconductor Devices Circuits and Systems **2010**.
[2] G. A. Rozgonyi, D. H. Hensler, Journal of Vac. Sci. & Tech. **1968**, 5, 194.
[3] M. M. Qazilbash, M. Brehm, B.-G. Chae, P. C. Ho, G. O. Andreev, B.-J. Kim, S. J. Yun, A. V. Balatsky, M. B. Maple, F. Keilmann, H.-T. Kim, D. N. Basov, Science **2007**, 318, 165108.
[4] M. D. Pickett, G. Medeiros-Ribeiro, R. S. Williams, Nat. Mat. **2012**, 12, 114.
[5] M. Nakano, K. Shibuya, D. Okuyama, T. Hatano, S. Ono, M. Kawasaki, Y. Iwasa, Y. Tokura, Nature **2012**, 487; J. Jeong, N. Aetukuri, T. Graf, T. D. Schladt, M. G. Samant, S. S. P. Parkin, Science **2013**, 339, 1402.
[6] B. Wu, A. Zimmers, H. Aubin, R. Ghosh, Y. Liu, R. Lopez, Phys. Rev. B **2011**, 84, 241410.
[7] G. Gopalakrishnan, D. Ruzmetov, S. Ramanathan, J. Mat. Sci. **2009**, 44, 5345; Z. Yang, S. Hart, C. Ko, A. Yacoby, S. Ramanathan, J of Appl. Phys. **2011**, 110, 033725.
[8] T. Driscoll, J. Quinn, M. Di Ventra, D. N. Basov, G. Seo, Y.-W. Lee, H.-T. Kim, D. R. Smith, Phys. Rev. B **2012**, 86, 094203.
[9] A. Zimmers, L. Aigouy, M. Mortier, M. Sharoni, S. Wang, K. G. West, J. G. Ramirez, I. K. Schuller, Phys. Rev. Lett. **2013**, 110, 056601; M. C. Bein, J. Mizsei, 2011 17th International Workshop on Thermal Investigations of ICs and Systems **2011**.
[10] D. Ruzmetov, S. Ramanathan, in *Thin Film Metal-Oxides*, (Ed: S. Ramanathan), Springer US, **2010**, 51.
[11] B.-J. Kim, Y. W. Lee, S. Choi, J.-W. Lim, S. J. Yun, H.-T. Kim, T.-J. Shin, H.-S. Yun, Phys. Rev. B **2008**, 77, 235401.
[12] C. N. Berglund, A. Jayaraman, Phys. Rev. **1969**, 185, 1034.
[13] M. Abbate, F. M. F. Degroot, J. C. Fuggle, Y. J. Ma, C. T. Chen, F. Sette, A. Fujimori, Y. Ueda, K. Kosuge, Phys. Rev. B **1991**, 43, 7263.
[14] L. A. Ladd, W. Paul, Solid State Comm. **1969**, 7, 425.
[15] A. Mansingh, R. Singh, S. B. Krupanidhi, Solid-State Elec. **1980**, 23, 649.
[16] J. P. Pouget, H. Launois, J. de Physique **1976**, 37, C4.
[17] M. Liu, H. Y. Hwang, H. Tao, A. C. Strikwerda, K. Fan, G. R. Keiser, A. J. Sternbach, K. G. West, S. Kittiwatanakul, J. Lu, S. A. Wolf, F. G. Omenetto, X. Zhang, K. A. Nelson, R. D. Averitt, Nature **2012**, 487, 1.
[18] J. Duchene, Journal of Solid State Chem. **1975**, 12, 303.
[19] L. Chua, S. Kang, Proc. IEEE **1976**, 64, 209.
[20] M. D. Pickett, R. S. Williams, Nanotech. **2012**, 23, 215202.
[21] F. A. Chudnovskii, L. L. Odynets, A. L. Pergament, G. B. Stefanovich, J. Solid State Chem. **1996**, 122, 95; C. R. Cho, S. I. Cho, S. Vadim, R. Jung, I. Yoo, Thin Solid Films **2006**, 495, 375.
[22] C. N. Berglund, IEEE Trans. Elec. Dev. **1969**, 16, 432.
[23] J. S. Lee, M. Ortolani, U. Schade, Y. J. Chang, T. W. Noh, Appl. Phys. Lett. **2007**, 90, 051907.
[24] G. Albright, J. Stump, C. P. Li, H. Kaplan, "Emissivity-corrected infrared thermal pulse measurement on microscopic semiconductor targets", presented at *Thermosense XXIII Conference*, Orlando, Fl, Apr 16-19, **2001**.

[25]     A. L. Kilcoyne, T. Tyliszczak, W. F. Steele, S. Fakra, P. Hitchcock, K. Franck, E. Anderson, B. Harteneck, E. G. Rightor, G. E. Mitchell, A. P. Hitchcock, L. Yang, T. Warwick, H. Ade, in *J. Synchrotron Radiat.*, Vol. 10, Denmark **2003**, 125.

[26]     F. M. F. de Groot, M. Grioni, J. C. Fuggle, J. Ghijsen, G. A. Sawatzky, H. Petersen, Phys. Rev. B **1989**, 40, 5715.

[27]     T. Yao, X. Zhang, Z. Sun, S. Liu, Y. Huang, Y. Xie, C. Wu, X. Yuan, W. Zhang, Z. Wu, G. Pan, F. Hu, L. Wu, Q. Liu, S. Wei, Phys. Rev. Lett. **2010**, 105, 226405.